\documentclass[12pt,aps,prb,preprint,showpacs,showkeys]{revtex4}  
\usepackage{amsmath}    
\usepackage{amsfonts}  
\usepackage{amssymb}
\usepackage{graphicx}   

\begin{document}

\title{Frequencies and energy levels of the weakly relativistic harmonic oscillator from the action variable}

\author{M.K.Balasubramanya$^{a)}$}

\affiliation{Department of Physical and Environmental Sciences\\
 Texas A\&M University-Corpus Christi\\
6300 Ocean Drive \\
 Corpus Christi, Texas, 78412}

\email{mirley.balasubramanya@tamucc.edu}

\date{\today}

\begin{abstract}
The frequency of a classical periodic system and the energy levels
of the corresponding quantum system can both be obtained using action
variables. We demonstrate the construction of two forms of the action
variable for a one dimensional harmonic oscillator in classical, relativistic
and quantum regimes. The relativistic effects are considered as perturbative,
within the context of a non-relativistic quantum formalism. The transition
of the relativistic quantum system to both classical relativistic
and classical non-relativistic regimes is illustrated in a unified
framework. Formulas for the frequency of a classical relativistic
oscillator and the energy eigenvalues of the corresponding quantum
oscillator for the weak relativistic case are derived. Also studied
are the non-relativistic and classical limits of these formulas which
provide valuable insights on the parallels between relativistic and
non-relativistic systems on the one hand and between classical and
quantum systems on the other.
\end{abstract}



\maketitle

\section{INTRODUCTION}

Classical periodic systems can be analyzed elegantly in terms of their
action and angle variables which constitute a set of canonically conjugate
momenta and coordinates. Action variables are proportional to $\oint p_{j}dq_{j}$,
where $(q_{j},p_{j})$ are the system's coordinates and canonical
momenta. For separable systems they are constants of motion in
the manner of angular momentum and energy. The frequencies of periodic
systems can be found using the functional relationship between the
action variables and total mechanical energy without requiring a complete
solution of the dynamical equations. The French astronomer and mathematician
Charles-Eug\`{e}ne Delaunay (1816-1872) invented action and angle
variables in the course of his study of periodicity of lunar motion
\cite{Delaunay}. These variables assumed importance during the early
days of quantum mechanics. Lord Rayleigh had shown that in a sinusoidally
oscillating system such as a pendulum whose string is shortened slowly,
the ratio of the energy to frequency, which is directly proportional
to the action variable, remains a constant. Following the language
of thermodynamics such motion was referred to as \char`\"{}adiabatic\char`\"{}
motion. At the first Solvay Conference in 1911, which considered the
issues that the new quantum ideas introduced into mechanics, it was
realized that the adaibatic invariance of the action variable in atomic
systems, in an environment of slowly varying electromagnetic fields,
would lead to atomic stability without transitions between states
\cite{Jammer}. Schwarzschild introduced into quantum theory the analytical
method of employing action variables \cite{Schwarzschild}. The quantization
rules of Sommerfeld, Wilson and Ishiwara required that the action
variables be integer multiples of $\hbar$ to account for the energy
spectra of atomic systems. \cite{Wilson,Sommerfeld,Ishiwara}. Ehrenfest
formulated the \char`\"{}adiabatic principle\char`\"{} according to
which a slow variation of some parameters of a periodic system's Hamiltonian
would result in a gradual change in the system's motion while maintaining
the constancy of the action variable \cite{Klein,Ehrenfest}. With
the establishment of the wave and matrix forms of quantum mechanics
the program of employing action variables in the quantum context did
not receive significant attention. In the JWKB approximation scheme
for the determination of bound quantum states the energy eigenvalues
are obtained by discretizing the action variable. Thus the quantum conditions
of the old quantum theory can be rigorously deduced as an approximate
result in the new quantum theory.

In 1983 Leacock and Padgett \cite{LPRL,LPR} presented a form of quantum
mechanics, patterned on the classical Hamilton-Jacobi theory and equivalent
to the Schrödinger theory, whose focus was a quantum version of the
action variable which reduces to the traditional action variable in
the classical limit. In this formalism the dynamical equation is a
postulated quantum Hamilton-Jacobi equation for the Hamilton's principal
function $S$ which generates a transformation from the coordinates and
momenta $(x_{i},p_{i})$ to angle and action variables $(w_{i},J_{i})$.
The bound states of a system are characterized by its quantum action
variables assuming values which are integral multiples of $\hbar$
in a natural way, and not through arbitrary imposition as was the
case in the Wilson-Sommerfeld scheme. Since the total mechanical energy
of the system is a function of its action variables it too assumes
discrete values for bound states. The examination of the dynamical
equations of this formalism of quantum theory, and the relation between
quantum action variables and energy, shows the classical-quantum correspondence
in a new light. A systematic development of canonical transformations
and the Hamilton-Jacobi theory in quantum mechanics is found in Ref.
11. Many systems admitting bound states have been studied
using this quantum Hamilton-Jacobi formalism \cite{Asiri,Bhalla}.
Our objectives here are two fold. We show, using a perturbative approach
based on two equivalent forms of the action variable, how a weakly
relativistic quantum oscillator can be treated within the non-relativistic
quantum Hamilton-Jacobi formalism to obtain its energy eigenvalues
that incorporate first order relativistic corrections. Two, we show
that the dynamics of this weakly relativitic oscillator has the correct
non-relativistic classical and non-relativistic quantum limits.

The symbols and notation we use for physical variables have the following
meaning. The angular frequency of a classical non-relativistic simple
harmonic oscillator of mass $m$ connected to a massless spring of
spring constant $k$ will be $\omega_0=\sqrt{\frac{k}{m}}$. The
suffix {\scriptsize $C$} refers to a classical variable; the absence
of this suffix indicates that the quantity being referred to is its
quantum counterpart. Similarly the suffix {\scriptsize $R$} refers
to a relativistic variable, and {\scriptsize $WR$} is the suffix
for a variable in the weak relativistic case.

\section{ACTION VARIABLE IN CLASSICAL MECHANICS}

\subsection{Classical Hamilton-Jacobi Theory}

The time evolution of a classical system is governed by its Hamiltonian
$H$ which is a function of its coordinates $x_{i}$, the conjugate
momenta $p_{i}$ and the time $t$. The dynamics of such a system
is determined by Hamilton's equations of motion \begin{eqnarray}
{\dot{x}}_{i}=\frac{\partial H(x_{i},p_{i},t)}{\partial p_{i}},\;\;\;\;
{\dot{p}}_{i}=-\frac{\partial H(x_{i},p_{i},t)}{\partial x_{i}}.
\label{hameq}
\end{eqnarray}
 The Hamiltonian of a particle of mass $m$ moving in one dimension
under the influence of a potential energy function $V(x)$ is given
by $H=\frac{p^{2}}{2m}+V(x)$. Such a time independent Hamiltonian
is a constant of the motion and is the total energy $E$ of the system.
Thus, \begin{eqnarray}
\frac{p^{2}}{2m}+V(x)=E.
\label{orbit}
\end{eqnarray}
Canonical transformations transform one set of
coordinate and momentum $(x,p)$ to another set $(X,P)$ while preserving
the form of Hamilton's equations. One such transformation is generated
by the function $W_{C}(x,P)$, whose arguments are the
\char`\"{}old\char`\"{} coordinate, $x$, and the \char`\"{}new\char`\"{}
momentum, $P$: 
\begin{eqnarray}
p=\frac{\partial W_{C}(x,P)}{\partial x},\;\; X=\frac{\partial W_{C}(x,P)}{\partial P}.
\label{trans}
\end{eqnarray}
 If this transformation transforms the Hamiltonian into a function
only of $P$, then, using ~(\ref{hameq}), 
\begin{eqnarray}
\dot{P}=-\frac{\partial H(P)}{\partial X}=0 & \Rightarrow\;\;\; P(t)=P,\;{\rm a\; constant},\nonumber \\
\dot{X}=\frac{\partial H(P)}{\partial P}=V_{0},\;\;{\rm a\; constant},\;\; & \Rightarrow\;\; X(t)=V_{0}\: t+X_{0}.
\label{transtwo}
\end{eqnarray}
Thus $X$ and $P$ evolve very simply in time; the former has a linear temporal progress
and the latter is a constant. The function $W_{C}(x,P)$, which
generates a canonical transformation in which the transformed Hamiltonian
is independent of the new coordinate $X$, is the Hamilton's {\it characteristic}
function. It is related to the Hamilton's {\it principal} function $S_{C}$
through $S_{C}(x,P,t)=W_{C}(x,P)-Et$, and, for the case of time independent
Hamiltonians, satisfies the Hamilton-Jacobi equation obtained by using
~(\ref{trans}) in ~(\ref{orbit}): 
\begin{eqnarray}
\frac{1}{2m}\left(\frac{\partial W_{C}(x,P)}{\partial x}\right)^{2}+V(x) & = & E(P).
\label{hj}
\end{eqnarray}

The use of this method to solve the dynamical problem involves the
following steps: (i) Define a suitable new constant momentum $P$,
(ii) Integrate Eq.~(\ref{hj}) to obtain $W_{C}(x,E(P))$, (iii)
Obtain $x(X,P)$ and $p(X,P)$ using Eq. ~(\ref{trans}), and (iv)
Express $X$ and $P$ in terms of the initial values $x_{0},p_{0}$
and $t$.

\subsection{Action-angle variables and periodic motion}

One particular form of Hamilton-Jacobi theory is especially suited
to the study of periodic motion. If an inspection of the Hamiltonian
indicates that the motion is periodic, then by a particular choice
of the new momentum $P$ we can evaluate the period of motion without
obtaining a complete solution of the dynamical problem. The new canonically
conjugate coordinate and momentum are chosen to be $X=w,\; P=J_{C}$
with \begin{eqnarray}
J_{C} & = & \frac{1}{2\pi}\oint p_{C}(x,E)dx,\label{oldj}\end{eqnarray}
 where $p_{C}(x,E)$, from ~(\ref{orbit}), is $\sqrt{2m[E-V(x)]}$
and the integral in phase space is performed over one cycle of the
periodic motion. $J_{C}$ is the classical action variable and $w$
the angle variable. We note that the integral for $J_{C}$
is the area enclosed in phase space by the path of the oscillator's
orbit. A new momentum, similar to $J_{C}$, will be defined in the
corresponding quantum formalism and will be referred to as $J$. Since
$J_{C}=J_{C}(E)$ we can invert it to obtain $E=E(J_{C})$. From Eq.~(\ref{transtwo})
the time evolution of the new coordinate is $w(t)=\omega\: t+w_{0}$
where the constant \char`\"{}velocity\char`\"{} is 
\begin{eqnarray}
\omega=\frac{\partial H(J_{c})}{\partial J_{C}} = \frac{\partial E(J_{c})}{\partial J_{C}}.
\label{wevolution}
\end{eqnarray}
It can be shown \cite{goldstein} that $\omega$ is the angular frequency
of this periodic motion. Thus the mathematical problem of finding
the frequency of motion for a periodic system
is reduced to that of performing the integral ~(\ref{oldj}), solving
for $E$ to get $E(J_{C})$, and evaluating $\partial E/\partial J_{C}$.
This is a simple and elegant method for evaluating the frequency of
a system known to be periodic. An equally simple method can be used
to obtain the energy eigenvalues of bound states in quantum mechanics.

An equivalent definition of $J_{C}$, which is useful for extending
the action variable into the quantum arena, is 
\begin{eqnarray}
J_{C} & = & \frac{1}{2\pi}\oint_{C}p_{C}(x,E)dx,
\label{compdef}
\end{eqnarray}
 where $p_{c}(x,E)$ is a complex valued function of the complex argument
$x$, and is defined as a suitable branch of 
\begin{eqnarray}
p_{c}(x,E) & = & \sqrt{2m\left[E-V(x)\right]}.
\label{pcdef}
\end{eqnarray}
The turning points $x_{1}$ and $x_{2}$ are defined by $p_{c}(x_{1},E)=p_{c}(x_{2},E)=0$.
These are also the branch points of $p_{c}(x,E)$ in the complex-$x$
plane. We choose a branch cut connecting $x_{1}$ and $x_{2}$ along
the real axis. $p_{c}(x,E)$ is chosen as that branch of the square
root which is positive along the bottom of the cut.
The counterclockwise rectangular contour $C$ wraps around this branch cut. The integral in ~(\ref{compdef}) is performed
by deforming the contour $C$ outward to the circular contour $\gamma$ which lies in an
annulus in which $p_{c}(x,E)$ is analytic, expanding $p_{c}(x,E)$
in a Laurent series in that annulus and integrating using
Cauchy's residue theorem. Sommerfeld was the first to employ this
contour integral technique in evaluating the action variable for the bound states 
of the electron in hydrogenic atoms.

An alternate construction of the action variable arises from another
canonical transformation scheme where the alternate Hamilton's principal
function $\tilde{S}_C(p,X,t)$ of the "old" momentum and "new" coordinate is used.
For conservative systems, the canonical transformation generated by the 
alternate Hamilton's characteristic function $\tilde{W}_C(p,X)$,
where $\tilde{S}_C(p,X,t)=\tilde{W}_C(p,X)-Et$, is $x = -\frac{\partial \tilde{W}_C(p,X)}
{\partial p}, P=-\frac{\tilde{W}_C(p,X)}{\partial X}$. The Hamilton-Jacobi equation
satisfied by this characteristic function is 
\begin{eqnarray}
H\left(-\frac{\partial \tilde{W}_C(p,X)}{\partial p},p\right)=E.
\label{HJforWpX}
\end{eqnarray}
As in the previous scheme, for periodic
systems, we can choose the new coordinate as the alternate action variable 
$J_C = -\frac{1}{2\pi}\oint_{C'}x_{C}(p,E)dp$,
where the clockwise rectangular contour $C'$ in the complex-$p$ plane 
surrounds the branch cut along the real axis connecting the turning momenta $p_{1}$ and $p_{2}$ 
and the sign of $x_{C}(p,E)$ is chosen positive below this cut. 
We will demonstrate the use of this alternate form of the action variable in both the
classical and quantum contexts for the weakly relativistic harmonic oscillator. The angle
variable, while important in describing the state of periodic motion, is not essential
for determining the frequency (and the energy levels in the quantum context) directly, and 
therefore will not be considered here.

\section{CLASSICAL WEAKLY RELATIVISTIC HARMONIC OSCILLATOR}

The relativistic motion of the harmonic oscillator is governed by
the Hamiltonian $H(x,p)=\sqrt{p^{2}c^{2}+m^{2}c^{4}}+\frac{1}{2}kx^{2}$.
The total mechanical energy of the relativistic oscillator will be
referred to as $E$, and $\tilde{E}=E-mc^{2}$ is its mechanical energy
in excess of its rest mass energy. The dimensionless energy related
parameter we will use is $\epsilon=\frac{\tilde{E}}{mc^{2}}$. We define the
weak relativistic motion of the oscillator as one characterized by $\epsilon<<1$, and
evaluate all dynamical variables up to the first order in $\epsilon$. We approximate 
the Hamiltonian for the relativistic oscillator by $H=\frac{p^{2}}{2m}-\frac{p^{4}}
{8m^{3}c^{2}}+\frac{kx^{2}}{2}$, retaining only the leading relativistic 
term in kinetic energy, and obtain the oscillator's approximate relativistic 
frequency using the two forms of the action variable discussed earlier. 
Such a study of the classical oscillator,
with an approximation of the relativistic Hamiltonian, although unnecessary
for frequency determination (at least three different series solutions exist for the frequency of a 
fully relativistic oscillator ~\cite{classreloscfreq}), is a necessary prelude to the consideration of the corresponding
quantum system within the framework of a non-relativistic quantum
theory, to be discussed in the next section. Using $T$ for the
non-relativistic kinetic energy $\frac{p^{2}}{2m}$ the correctional
relativistic term in $H$ can be written as $-\frac{1}{2}T(\frac{T}{mc^{2}})$.
The factor $\frac{T}{mc^{2}}$ is of order $\epsilon$, and can be
treated as a perturbative parameter in the non-relativistic Hamiltonian.
This approximation of the Hamiltonian is good for studying motion
in the weakly relativistic case. For consistency we should retain
terms only up to order $\epsilon^{1}$ in expressions for $p_{CWR}$,
$J_{CWR}$ and ${\omega}_{WR}$.

We first consider the classical action variable in the $\oint pdx$
form. The classical orbit equation is $\frac{{p_{CWR}}^{2}}{2m}-\frac{{p_{CWR}}^{4}}
{8m^{3}c^{2}}+\frac{kx^{2}}{2}=\tilde{E}$. Extending this into the complex-$x$ plane 
the momentum is a branch of 
\begin{equation}
p_{CWR}(x,\tilde{E})=\sqrt{2} \; mc\:\sqrt{1-\sqrt{1-\frac{2}{mc^{2}}
\left(\tilde{E}-\frac{1}{2}kx^{2}\right)}}.
\label{wrorbiteqn}
\end{equation}
Two turning points emerge at $x_{1}$ and $x_{2}$, identical in form
to those in the fully relativistic
case, given by $-x_{1} = x_{2} = \sqrt{\frac{2\tilde{E}}{k}}$. These are also 
two of the branch points of $p_{CWR}(x,E)$. There are two additional branch points, $x_{3WR}$
and $x_{4WR}$, arising from the weak relativistic correction in the Hamiltonian, given by 
\begin{equation}
-x_{3WR}=x_{4WR}=\sqrt{\frac{2\tilde{E}}{k}}\sqrt{1-\frac{1}{2\epsilon}}.
\label{weakreltp}
\end{equation}
For $\epsilon<<\frac{1}{2}$ these branch points are on the imaginary axis with $x_{4WR}$ farther from the origin than $x_2$.
We choose one branch cut of $p_{CWR}$ to connect $x_{1}$ and $x_{2}$ along the real axis, and two other cuts, 
each connecting $x_{3WR}$ and $x_{4WR}$ to $x = \infty$ along the imaginary axis.  We choose
the sign of $p_{CWR}(x,\tilde{E})$ to be positive below the cut joining $x_{1}$ and $x_{2}$.
$p_{CWR}$ is analytic for $x_2 < |x| < x_{4WR}$ and its Laurent series in this annulus can be written as 
\begin{equation}
p_{CWR}(x,E)=i\sqrt{mk}\sqrt{1+\epsilon} \; x\left[1-\left(\frac{x_{2}}{x}\right)^{2}\right]^{\frac{1}{2}}
\sqrt{\frac{2}{1+\sqrt{1-(\frac{x}{x_{4WR}})^{2}}\sqrt{1-2\epsilon}}} =\sum_{j=-\infty}^{\infty}A_{Wj}x^{3-2j}.
\label{weakpwithepsilon}
\end{equation}
Here the square roots can be expanded using binomial series. The
action variable $J_{CWR}(\tilde{E})$, defined as 
$\frac{1}{2\pi}\oint_{C_{WR}}p_{CWR}(x,E)dx$,
where $C_{WR}$ is a counterclockwise rectangular contour that hugs the branch cut 
connecting $x_1$ and $x_2$,
is evaluated up to order $\epsilon^{1}$ by expanding the square roots
in Eq.~(\ref{weakpwithepsilon}) and integrating using Cauchy's residue theorem. That yields 
\begin{equation}
J_{CWR}(\tilde{E}) = \frac{1}{2\pi}(2\pi i)A_{W2}
\approx i\; . \; i\sqrt{mk}\;\left(-\:
\frac{\tilde{E}}{m{{\omega}_{0}}^{2}}\right)\;
\left[1+\frac{3}{16}\frac{\tilde{E}}{mc^{2}}\right]\;=\;
\frac{\tilde{E}}{{\omega}_{0}}\;\left(1+\frac{3}{16}\epsilon\right).
\label{weakjcrpdx}
\end{equation}
This differs from $\frac{\tilde{E}}{{\omega}_{0}}$, the action variable
for the non-relativistic harmonic oscillator, with a correction of order $\epsilon$. 
The angular frequency is obtained from 
\begin{equation}
\frac{1}{{\omega}_{WR}}=\frac{\partial J_{CWR}}{\partial E}=\frac{1}{{\omega}_{0}}(1+\frac{3}{8}\epsilon),
\label{omegawrpdx}
\end{equation}
which shows that the weak relativistic correction results in a fractional decrease of about $\frac{3}{8}\epsilon$ 
in the oscillator's frequency.

We now consider the second form of the action variable for the weakly
relativistic case. The orbit equation can be solved for the coordinate
to yield $x_{CWR}(p,\tilde{E})=\sqrt{\frac{2}{k}}\:\left[\tilde{E}-
\frac{p^{2}}{2m}+\frac{p^{4}}{8m^{3}c^{2}}\right]^{\frac{1}{2}}$.
The two physical turning momenta of this oscillator, where the coordinate vanishes, are defined by
\begin{equation}
-p_{WR1}=p_{WR2}=\sqrt{2}mc\left[1-\left(1-2\epsilon\right)^{\frac{1}{2}}\right]^{\frac{1}{2}}\;
\approx\;\sqrt{2m\tilde{E}}\left(1+\frac{\epsilon}{4}\right).
\label{turnmomwr}
\end{equation}
 These are very nearly the non-relativistic turning momenta, given by 
$-p_{1} = p_{2} = \sqrt{2m\tilde{E}}$. They contain a
weak relativistic correction of order $\epsilon$. There are two additional
branch points, $p_{WR3,4}$ of (weak) relativistic origin given by 
\begin{equation}
-p_{WR3}=p_{WR4}=\sqrt{2}mc\left[1-\left(1-2\epsilon\right)^{\frac{1}{2}}\right]^{\frac{1}{2}}.
\label{wrxrelbranchpt}
\end{equation}
 
We write $x_{CWR}(p,\tilde{E})$ in the form 
\begin{eqnarray*}
x_{CWR}(p,\tilde{E})=\sum_{j=-\infty}^{\infty}A'_{Wj}p^{3-2j}=\frac{-i}{\sqrt{mk}}
\left[\frac{1}{\sqrt{2}}+\frac{1}{\sqrt{2}}\left(1-2\epsilon\right)^{\frac{1}{2}}\right]^{\frac{1}{2}} 
p\left[1-\left(\frac{p_{WR2}}{p}\right)^{2}\right]^{\frac{1}{2}}
\left[1-\left(\frac{p}{p_{WR4}}\right)^{2}\right]^{\frac{1}{2}}.
\end{eqnarray*}
 Using this we extend $x_{CWR}(p,\tilde{E})$ into the complex-$p$
plane. We choose one branch cut of $x_{CWR}(p,\tilde{E})$ from $p_{1WR}$
to $p_{2WR}$ along the real axis (the function is chosen positive just below this cut),
and two other cuts, each of which joins $p_{3WR}$ and $p_{4WR}$ to $p=\infty$ along the real axis.
The alternate definition of $J_{CWR}(\tilde{E})$ is 
\begin{equation}
J_{CWR}(\tilde{E})=-\:\frac{1}{2\pi}\oint_{C_{WR}'}x_{CWR}(p,\tilde{E})dp,
\label{altjcwr}
\end{equation}
 where the counterclockwise rectangular contour $C'_{WR}$ wraps around the branch cut connecting
$p_{1WR}$ and $p_{2WR}$. For $p_{WR2}<|p|<p_{WR4}$ we expand $x_{CWR}$
in a Laurent series and evaluate $J_{CWR}$ to obtain 
\begin{equation}
J_{CWR}=\frac{\tilde{E}}{{\omega}_{0}}\;\left[\frac{2}{1+\sqrt{1-2\epsilon}}\right]^{\frac{1}{2}}\;
\left[1-\frac{1}{8}\left(\frac{p_{WR2}}{p_{WR4}}\right)^{2}\ldots\right].
\label{altjcwreval}
\end{equation}
This representation of the action variable contains a series in powers
of $(p_{WR2}/p_{WR4})^{2} = \left(\frac{1-\sqrt{1-2\epsilon}}
{1+\sqrt{1-2\epsilon}}\right) \approx \frac{\epsilon}{2}$. To order $\epsilon$
it is $\frac{\tilde{E}}{{\omega}_{0}}[1+\frac{3}{16}\epsilon]$, consistent
with ~(\ref{weakjcrpdx}). The angular frequency is found from 
\begin{equation}
\frac{1}{{\omega}_{WR}}=\frac{dJ_{CWR}}{d\tilde{E}}=\frac{1}{{\omega}_{0}}\;
\frac{d}{d\epsilon}\;\left[\epsilon\left\{ \frac{2}{1+\sqrt{1-2\epsilon}}
\right\} ^{\frac{1}{2}}\left\{ 1-\frac{1}{8}\left(\frac{1-\sqrt{1-2\epsilon}}
{1+\sqrt{1-2\epsilon}}\right)\ldots\right\} \right]\approx\frac{1}
{{\omega}_{0}}(1+\frac{3}{8}\epsilon),
\label{omegawrxdp}
\end{equation}
which agrees with the result in Eq.~(\ref{omegawrpdx}). Further both results agree, 
up to order $\epsilon^1$, with the expressions for the angular frequency obtained by using the fully 
relativistic Hamiltonian ~\cite{classreloscfreq}. 

\section{HARMONIC OSCILLATOR IN QUANTUM HAMILTON-JACOBI THEORY}

\subsection{Quantum Action Variable}

We summarize here the formalism of Hamilton-Jacobi quantum mechanics
\cite{LPRL,LPR}, equivalent to other better known ones, and apply
it to the special case of 1-D one particle systems with Hamiltonians
of the form $\hat{H}=\frac{{\hat{p}}^{2}}{2m}+\hat{V}({\hat{x}})$.
The measurable values of the observables $\hat{H},\hat{p}$ and $\hat{x}$
are their eigenvalues $E$, $p$ and $x$ respectively. The equations
of quantum canonical transformation are written in terms of the eigenvalues
and functions of eigenvalues of these observables. Using the quantum
characteristic function $W(x,P)$ these transformation equations,
equivalent to the classical ones in Eq.~(\ref{trans}), are 
\begin{eqnarray}
p=\frac{\partial W(x,P)}{\partial x},\;\;\;\;\;\; X=\frac{\partial W(x,P)}
{\partial P}.
\label{qcanontrans}
\end{eqnarray}
 The \textit{quantum Hamilton-Jacobi equation}, for systems with time
independent Hamiltonians, is 
\begin{equation}
-i\hbar\frac{{\partial}^{2}W(x,E(P))}{\partial x^{2}}+\left(\frac{\partial W(x,E(P))}
{\partial x}\right)^{2}=E(P)-V(x).
\label{qhj}
\end{equation}
 This can be either postulated or derived from Schrödinger's equation
for the state $\psi(x,t)$, which is written in the form $e^{\frac{i}{\hbar}S(x,t)}$,
where $S(x,t)$ is Hamilton's principal function in the quantum context. 
For systems with time independent Hamiltonians, $W$ and $S$ are
related through $S(x,P,t)=W(x,P)-Et$. Physical boundary conditions
have to be imposed on $W(x,E(P))$ to complete its definition. We
note that this equation resembles the classical Hamilton-Jacobi equation
~(\ref{hj}), except for the additional term involving $\hbar$,
and reduces to it in the limit $\hbar\;\rightarrow\;0$. The dynamics
described by this equation is non-relativistic and is equivalent in all respects
to the Schrödinger formalism of quantum mechanics. We can use such an
equation only in a perturbative sense to treat systems with weak relativistic
terms in their Hamiltonian.

$p(x,E)=\frac{\partial W(x,P(E))}{\partial x}$ is the quantum analog
of the classical momentum function $p_{C}(x,E)$ and, following Ref.
10, will be referred to as the \textit{quantum momentum
function}. Using this definition of $p(x,E)$ in Eq. ~(\ref{qhj}) we obtain 
\begin{eqnarray}
-i\hbar\frac{\partial p(x,E)}{\partial x}+p^{2}(x,E)\;=
\;2m[E-V(x)]\;=\;{p_{c}}^{2}(x,E).
\label{peq}
\end{eqnarray}
 We note that the square of this quantum momentum function reduces to the square of $\pm p_{c}(x,E)$, the
classical momentum function in the phase space orbit equation, in the limit $\hbar\;\rightarrow\;0$.
The physical boundary condition on $p(x,E)$ is that for all $x$
\begin{equation}
\lim_{\hbar\rightarrow0}p(x,E)\;\;=\;\;+p_{c}(x,E).
\label{claslim}
\end{equation}
 Eq.~(\ref{peq}), along with ~(\ref{claslim}), defines $p(x,E)$
which we will use in the construction of the quantum action variable
$J$. Equation ~(\ref{peq}) is the Ricatti (nonlinear) form of the Schrödinger (linear) differential
equation. Extending the wave function $\psi_E(x,t) = \phi(x,E)e^{-\frac{iEt}{\hbar}}$ into the
complex $x$ plane it is easily shown that the nodes of $\phi(x,E)$ are also the poles 
of $p(x,E)$. Using oscillation theorems for the linear differential
equation it can be shown that (i) $p(x,E)$ has simple poles of residue $-i\hbar$ on the
real axis between the two turning points independent of the system's energy, and
(ii) there are fixed poles whose locations and residues are determined by the specific nature of 
the potential.\cite{LPR,Bhalla}. The number
of poles between the turning points counts the level of excitation of this quantum system.

Further development of this quantum formalism for periodic systems,
where action-angle variables can be employed, requires the specification
of a new momentum $P$, which is the \textit{quantum action variable}
$J$. Following the definition of the classical action variable it
is defined as the contour integral in the complex $x$ plane, 
\begin{equation}
J(E)=\frac{1}{2\pi}\oint_{C}p(x,E)dx,
\label{qj}
\end{equation}
 with the counterclockwise rectangular contour $C$ tightly wrapping the real axis between the two physical turning
points of the {\it classical} momentum function $p_{C}(x,E) = \sqrt{E-V(x)}$. Since 
$C$ encloses only a finite number of poles of $p(x,E)$, deforming $C$
to encircle the poles and integrating using the residue theorem leads
to the discretization of $J$: 
\begin{eqnarray}
J=\;\frac{1}{2\pi}(2\pi i)\{ n(-i\hbar)\}\;\;=\;\; n\hbar.
\label{Jquant}
\end{eqnarray}
 This is in contrast to the classical action variable which assumes
continuous values. At first glance, Eq.~(\ref{Jquant}) seems to be the 
Wilson-Sommerfeld quantization condition or the JWKB quantum rule, but
it is not. The action variable (or the Sommerfeld phase integral) used
in those conditions is the {\it classical} action variable; the one used here
is the quantum action variable whose definition is based on the quantum
momentum function $p(x,E)$ which is a solution of Eq.~(\ref{peq}) whose 
basis is in quantum mechanics. Finally, By deforming $C$ outward and integrating we 
capture the energy dependence of the quantum action variable and obtain
$J(E)$. Thus $J(E)=n\hbar$, and inverting this, we obtain the system's energy 
eigenvalues, $E=E(J=n\hbar)$.

As will be shown later, the integral in Eq.~(\ref{qj}) can be performed
without obtaining a solution of Eq.~(\ref{peq}) all over the complex
$x$ plane. The quantum energy eigenvalues of a system that is classically
periodic can thus be obtained more simply by using the discretizatized
nature of the quantum action variable than by imposing boundary conditions on the wave function.
It can be shown that the poles of $p(x,E)$ on the real axis between
the physical turning points coalesce in the classical limit, and
form the branch cut of $p_{c}(x,E)$. The mechanism of this pole coalescing
is demonstrated in Ref. 10. The residue of a pole of $p(x,\tilde{E})$
between the turning points is proportional to $\hbar$, and so is
the spacing between neighboring poles. In the limit $\hbar\rightarrow0$,
for a fixed energy of the oscillator, the real axis between the turning
points is riddled with poles, with the quantum momentum function having
opposite signs on the sides of the pole above and below the real axis,
giving rise to the branch cut of $p_{C}(x,E)$.

We will construct two such equivalent forms of the action variable
$J$ for the simple harmonic oscillator in the quantum context and
demonstrate that both forms of $J$ are discrete. Since the energy
is a function of the action variable, it too is discrete rather than continuous. It
is important to recognize that (1) this quantum action variable differs
from the classical action variable employed by Sommerfeld and others
for quantization, (2) the quantum action variable naturally assumes
values that are integral multiples of $\hbar$ and (3) the energy eigenvalues
obtained by evaluating the action variable are exact. By making appropriate
approximations of this action variable as a function of energy we
obtain approximate energy eigenvalues.

\subsection{$\oint pdx$ form of quantum action variable}

The quantum action variable for the harmonic oscillator is defined
through Eq.~(\ref{qj}) where the rectangular counterclockwise contour $C$ closely wraps around the real axis
between the turning points $x_1$ and $x_2$.
For evaluating the integral we deform $C$ outward into the circular contour $\gamma$ centered at the origin, and use
the Laurent series for $p(x,E)$ in an origin centered annulus that
includes $\gamma$. The boundary condition on $p(x,E)$ indicates
the form of the Laurent series that it has in this annulus. The Laurent series
for the classical momentum function is
\begin{equation}
p_C(x,E)=\sqrt{E-\frac{1}{2}kx^2}= i\sqrt{mk} x \left[  1 -  
\left(\frac{x_2}{x}\right)^2 \right]^{\frac{1}{2}}= \sum_{j=1}^{\infty}a_{j}x^{3-2j}.
\label{pdxseries}
\end{equation}
We choose the form for the series for the quantum momentum function that is similar
to that of the series for $p_{C}(x,E)$ and write it as 
\begin{equation}
p(x,E)=\sum_{j=1}^{\infty}b_{j}x^{3-2j}.
\label{pdxseries}
\end{equation}
 Substituting this in Eq.~(\ref{peq}) and equating coefficients
of like powers of $x$ we get $b_{1}=\pm i\sqrt{mk}$. We choose the $+$ sign
here since $a_1 = +i\sqrt{mk}$; the boundary condition on $p(x,E)$ requires
$b_j \rightarrow a_j$ in the classical limit.
The next coefficient is $b_{2}=-i\sqrt{\frac{m}{k}}E+i\frac{\hbar}{2}$.
Using the residue theorem we get 
\begin{equation}
J(E)=2\pi i\:(\frac{1}{2\pi})\: b_{2}=E\sqrt{\frac{m}{k}}-\frac{\hbar}{2}
\label{jpdxeval}
\end{equation}
 This, along with the discretization condition in Eq.~(\ref{Jquant}),
yields $E=(n+\frac{1}{2})\hbar$, the energy eigenvalues of the harmonic
oscillator.

\subsection{Equivalent quantum Hamilton-Jacobi equation}

We present here a new version of the quantum Hamilton-Jacobi equation that is equivalent to
the one in Eq.~(\ref{qhj}) using an alternate quantum canonical
transformation scheme. While the dynamical equation can be postulated
for the quantum generating function $\tilde{S}(p,X,t)$ that generates
the transformation \begin{eqnarray*}
x=-\frac{\partial\tilde{S}(p,X,t)}{\partial p},\;\;\; P=-\frac{\partial\tilde{S}(p,X,t)}{\partial X},\end{eqnarray*}
 and show its equivalence to the Schrödinger equation, we choose the
reverse route and derive it from the latter. Starting with the Schrödinger
equation 
\begin{equation}
\left[\frac{{\hat{p}}^{2}}{2m}+\hat{V}(\hat{x})\right]|\phi>\;=\; i\hbar\frac{\partial|\phi>}{\partial t}
\label{Schrodinger}
\end{equation}
 and using the momentum representation where $\hat{x}\rightarrow i\hbar\frac{\partial}{\partial p},
\;\hat{p}\rightarrow p$ and $|\phi>\rightarrow\phi(p,E,t)$, we get 
\begin{equation}
\left[\frac{p^{2}}{2m}+V\left(i\hbar\frac{\partial}{\partial p}\right)\right]\phi=
i\hbar\frac{\partial\phi}{\partial t}.
\label{pSchrodinger}
\end{equation}
 Introducing the alternate quantum Hamilton's principle function $\tilde{S}(p,X,t)$
through $\phi(p,E(X),t)=e^{\frac{i}{\hbar}\tilde{S}(p,X,t)}$ and
defining $G$ from 
\begin{equation}
V(i\hbar\frac{\partial}{\partial p})\:\phi(p,E,t)=G\left(\hbar,\tilde{S}_{p},\tilde{S}_{pp},\ldots\right)\: 
e^{\frac{i}{\hbar}\tilde{S}(p,X,t)}
\label{Gdefinition}
\end{equation}
 we obtain 
\begin{equation}
\frac{p^{2}}{2m}+G(\hbar,\tilde{S}_{p},\tilde{S}_{pp},\ldots)=-\tilde{S}_{t}.
\label{phiequation}
\end{equation}
 Here suffixes denote partial differentiation. We define the alternate
quantum characteristic function $\tilde{W}(p,X)$ through the relation
$\tilde{S}(p,X,t)=\tilde{W}(p,X)-Et$ and use it in Eq.~(\ref{phiequation})
to get 
\begin{equation}
\frac{p^{2}}{2m}+G(\hbar,\tilde{W}_{p},\tilde{W}_{pp},\ldots)=E.
\label{WHamiltonJacobi}
\end{equation}
 This is the quantum Hamilton-Jacobi equation for $\tilde{W}(p,X)$
which generates a canonical transformation from $(x,p)$ to $(X,P)$.
Use of the quantum canonical transformation equation,
$x=-\frac{\partial\tilde{W}}{\partial p}$, in Eq.~(\ref{WHamiltonJacobi})
leads to 
\begin{equation}
\frac{p^{2}}{2m}+G\left(\hbar,x(p,E),\frac{\partial x(p,E)}{\partial p},\ldots\right)=E.
\label{alternateHamiltonJacobi}
\end{equation}
 This is the equivalent of Eq.~(\ref{peq}) under this alternate
canonical transformation scheme. The physical boundary condition on
$x(p,E)$ is that in the limit $\hbar\rightarrow 0$, it 
should reduce to the classical coordinate function, $x_{C}(p,E)$. 
Specializing to the case of the simple harmonic oscillator we get
\begin{equation}
G=\frac{i}{\hbar}\frac{\partial x(p,E)}{\partial p}
-\frac{1}{{\hbar}^{2}}{x^{2}}(p,E).
\label{Gequation}
\end{equation}
 Substituting this in Eq.~(\ref{alternateHamiltonJacobi}) we obtain
the quantum equivalent of the classical orbit equation: 
\begin{equation}
i\hbar\frac{\partial x}{\partial p}+x^{2}
=\frac{2}{k}\left[E-\frac{p^{2}}{2m}\right].
\label{xpquantumorbit}
\end{equation}
 In the limit $\hbar\rightarrow 0$ this equation becomes, for real
values of $x$ and $p$, the classical orbit equation that constrains
the harmonic oscillator to an elliptical path in phase space.

We choose that quantum canonical transformation which makes the quantum
action variable $J$ the new coordinate. It is defined as the contour
integral 
\begin{equation}
J=-\frac{1}{2\pi}\oint_{C'}x(p,E)dp
\label{xdpJ}
\end{equation}
 in the complex-$p$ plane, with the clockwise rectangular contour $C'$ enclosing
the real axis between the two turning momenta $p_{1}$ and $p_{2}$.
A comparison of Eq.~(\ref{xpquantumorbit}) with Eq.~(\ref{peq})
shows that, as in the case of the quantum momentum function $p(x,E)$,
the quantum coordinate function $x(p,E)$ has a finite number of poles
on the real axis between $p_{1}$ and $p_{2}$, each of residue $+i\hbar$. 
In the classical limit these poles coalesce to form the
branch cut of $x_{C}(p,E)$. The steps for the determination of energy
eigenvalues here parallels that described in the previous section.
We evaluate $J$ in two ways, one by outward deformation of $C'$ into the origin 
centered circular contour ${\gamma}'$ and integrating
to obtain $J=J(E)$, and, two, through inward deformation, encircling
the poles and integrating, to get $J=n\hbar$.

For $|p|>p_{2}$, the form of the Laurent series for the classical momentum function 
$x_C(p,\tilde{E})$ is  
\begin{eqnarray*}
x_{C}(p,\tilde{E})=\sqrt{\frac{2}{k}\left[E - \frac{p^2}{2m}\right]} = 
\frac{-i}{\sqrt{mk}} p \left[ 1 - \left(\frac{p_2}{p}\right)^2  \right]^{\frac{1}{2}} = 
\sum_{j=-\infty}^{\infty}a'_{j}p^{3-2j}.
\end{eqnarray*}
We use a similar series form for the quantum coordinate
function $x(p,E)$ valid for this region of the complex plane:
\begin{equation}
x(p,E)=\sum_{j=1}^{\infty}b'_{j}p^{3-2j}.
\label{xexpansion}
\end{equation}
Substituting this in Eq.~(\ref{alternateHamiltonJacobi}),
and imposing the physical boundary condition on $x(p,E)$, we obtain
\begin{equation}
b'_{1}=\frac{-i}{\sqrt{mk}},\; b'_{2}=-i\left(\frac{\hbar}{2}\;-\;
\frac{E}{{\omega}_{0}}\right)
\label{bcoefficients}
\end{equation}
 Expanding the contour $C'$ outward to $\gamma'$ and evaluating
$J$ we get 
\begin{equation}
J=-\:\frac{1}{2\pi}\:(2\pi i)\: b'_{2}=\frac{E}{\omega_{0}}-\frac{\hbar}{2}.
\label{jqxdpeval}
\end{equation}
 Shrinking the contour inward to encircle the poles on the real axis
between $p_{1}$ and $p_{2}$ and integrating yields $J=n\hbar$.
So, $\frac{E}{\omega_{0}}-\frac{\hbar}{2}=n\hbar$, and inverting
this relation we obtain the energy eigenvalues of the simple harmonic
oscillator, $E=(n+\frac{1}{2})\hbar\omega_{0}$.

This alternate form of the quantum action variable has the same physical
content as the previous form. While the first form is easier to evaluate
for systems whose potential energy functions are not necessarily quadratic
in $x$, the second form is better suited for the weakly relativistic
oscillator, as will be shown in the next section.

\section{QUANTUM MECHANICS OF WEAKLY RELATIVISTIC OSCILLATOR}

\subsection{Integral $\oint pdx$ form of action variable}

The Hamiltonian for the weakly relativistic oscillator is $\frac{{\hat{p}}^{2}}{2m}
-\frac{{\hat{p}}^{4}}{8m^{3}c^{2}}+\frac{1}{2}k{\hat{x}}^{2}$.
Using this in the quantum Hamilton-Jacobi equation for $W(x,P)$ (we
omit the subscript $WR$ for $W$), which is obtained
from Schrödinger's equation following the steps outlined in Section
IV.A, we get 
\begin{eqnarray}
\tilde{E}=\frac{1}{2}kx^{2}+\frac{1}{2m}\left(\frac{\partial W}{\partial x}
\right)^{2}-\frac{i\hbar}{2m}\frac{{\partial}^{2}W}{\partial{x}^{2}}-\frac{1}
{8m^{3}c^{2}}\left[\left(\frac{\partial W}{\partial x}\right)^{4}-6i\hbar\left(\frac{\partial W}
{\partial x}\right)^{2}\frac{{\partial}^{2}W}{\partial x^{2}}-\right.\nonumber \\
{\hbar}^{2}\left\{ 4\frac{\partial W}{\partial x}\frac{{\partial}^{3}W}{\partial x^{3}}+
3\left(\frac{{\partial}^{2}W}{\partial x^{2}}\right)^{2}\right\} +
\left.i{\hbar}^{3}\frac{{\partial}^{4}W}{\partial x^{4}}\right]
\label{HJWpdx}
\end{eqnarray}
 Using the quantum canonical transformation equation $p=\partial W/\partial x$
in Eq.~(\ref{HJWpdx}) results in
\begin{eqnarray}
\tilde{E}-\left[\frac{p^{2}}{2m}+\frac{1}{2}kx^{2}-\frac{p^{4}}{8m^{3}c^{2}}\right]=
\;\;\;\;\;\;\;\;\;\;\;\;\;\;\;\;\;\;\;\;\;\;\;\;\;\;\;\;\;\;\;\;\;\;\nonumber \\
-\frac{i\hbar}{2m}\frac{\partial p}{\partial x}+\frac{\hbar}{2m^{3}c^{2}}
\left[\frac{3}{2}ip^{2}\frac{\partial p}{\partial x}+\hbar\left\{ \frac{3}{4}
\left(\frac{\partial p}{\partial x}\right)^{2}+p\frac{{\partial}^{2}p}{\partial x^{2}}\right\} 
-\frac{i{\hbar}^{2}}{4}\frac{{\partial}^{3}p}{\partial x^{3}}\right]
\label{weakpdxmom}
\end{eqnarray}
Eq.~(\ref{weakpdxmom}) shows the presence of terms proportional
to $\hbar$ or its higher powers, which are absent in the corresponding
classical orbit equation. Further, in the non-relativistic limit it
reduces to Eq.~(\ref{peq}).

Using the solution $p_{WR}$ of Eq.~(\ref{weakpdxmom}) satisfying
the physical boundary condition (i.e., has the correct classical limit) we define the
quantum action variable as 
\begin{equation}
J_{WR}=\frac{1}{2\pi}\oint_{C_{WR}}p_{WR}(x,\tilde{E})dx,
\label{jquantumwrdef}
\end{equation}
 using the same contour $C_{WR}$ as in the classical case.
Following the form of $p_{CWR}(x,\tilde{E})$ in Eq.~(\ref{weakpwithepsilon})
the Laurent series for $p_{WR}(x,\tilde{E})$ in the annulus $x_{2}<|x|<x_{4WR}$, 
correct to order $\epsilon^{1}$, is of the form 
\begin{equation}
p_{WR}(x,\tilde{E})=\left[\sum_{j=1}^{\infty}b_{j}x^{3-2j}\right]
\left[1+\frac{\epsilon}{4}\left\{ B_{0}-B_{1}\left(\frac{x}{x_{2}}\right)^{2}\right\} \right].
\label{wrpquantumseries}
\end{equation}
 The non-relativistic quantum momentum function in Eq.~(\ref{pdxseries})
has been modified here by the addition of a relativistic term of
order $\epsilon$. We have chosen the form of this series such that
(i) the relativistic modification is of the first order in $\epsilon$
and in the limit $\epsilon \rightarrow 0$ we get $p_{WR}(x,\tilde{E})\rightarrow p(x,\tilde{E})$,
and (ii) in the limit $\hbar \rightarrow 0$ we obtain $p_{WR}(x,\tilde{E})\rightarrow p_{CWR}(x,\tilde{E})$.
The coefficients $b_{j}$ are known from the non-relativistic quantum
case previously considered (see Section IV.B). Substituting this series
form of the solution in Eq.~(\ref{weakpdxmom}) we obtain the coefficients
$B_{0}$ and $B_{1}$: 
\begin{equation}
B_{0}=1,\;\; B_{1}=1+\frac{7\hbar{\omega}_{0}}{4\tilde{E}}.
\label{Bcoefficients}
\end{equation}
 The coefficient of $x^{-1}$ in $p_{WR}(x,\tilde{E})$ is $im{\omega}_{0}\; x_{2}^{2}\;(b_{1}B_{0}-b_{0}B_{1})$.
Evaluating $J$ by deforming the contour $C_{WR}$ outward to the origin centered circular contour ${\gamma}_{WR}$ previously considered,
we get \begin{eqnarray}
J_{WR}\;\;=\;\;2\pi i\:\left(\frac{1}{2\pi}\right)\: im{\omega}_{0}\; x_{2}^{2}\;(b_{1}B_{0}-b_{0}B_{1})\nonumber \\
=\frac{\tilde{E}}{{\omega}_{0}}\;\left[1+\epsilon\left\{ \frac{3}{16}+\frac{7}{16}
\left(\frac{\hbar{\omega}_{0}}{\tilde{E}}\right)-\frac{17}{64}
\left(\frac{\hbar{\omega}_{0}}{\tilde{E}}\right)^{2}\right\} \right]-\frac{\hbar}{2}
\label{jwrpdxeval}
\end{eqnarray}

A comparison with Eq.~(\ref{jpdxeval}) shows the relativistic corrections
present in this quantum action variable up to order $\epsilon$. Solving
Eq.~(\ref{jwrpdxeval}) for $\tilde{E}$, we get 
\begin{equation}
\tilde{E}\;=\;\tilde{E}(J_{WR}=n\hbar)\;\;=\;\;\left[\left(n+\frac{1}{2}\right)-\frac{3}{16}
\left\{ \left(n+\frac{5}{3}\right)^{2}-\frac{25}{9}\right\} 
\frac{\hbar{\omega}_{0}}{mc^{2}}\right]\;\hbar{\omega}_{0}
\label{wrpdxlevels}
\end{equation}
 This gives energy eigenvalues with a first order relativistic correction
that is proportional to the dimensionless energy parameter $\frac{\hbar\omega_{0}}{mc^{2}}$,
which measures the level separation in the non-relativistic case.
The energy levels in the weak relativistic case are lower than in
the non-relativistic case, and the lowering is predominantly quadratic
in the quantum number $n$. The seperation between energy levels is proportional to $n$.

\subsection{$-\oint xdp$ form of action variable}

We now apply the formalism developed in Section IV.C to construct the
quantum action variable in its alternate form. Using the Hamiltonian
for the weakly relativistic oscillator in the quantum Hamilton-Jacobi
equation for the alternate characteristic function $\tilde{W}(p,X)$ we get 
\begin{equation}
-i\hbar\frac{{\partial}^{2}\tilde{W}}{\partial p^{2}}+\left(\frac{\partial\tilde{W}}
{\partial p}\right)^{2}=\frac{2}{k}\left[\tilde{E}-\left(\frac{p^{2}}{2m}
-\frac{p^{4}}{8m^{3}c^{2}}\right)\right].
\label{HJweakqRxdP}
\end{equation}
 In comparison to the quantum Hamilton-Jacobi equation ~(\ref{HJWpdx}),
this alternate form is simpler as it has a single quantum term, and is to be preferred in the study
of this oscillator. Introducing the quantum canonical transformation
equation $x=-\partial\tilde{W}/\partial p$ in Eq.~(\ref{HJweakqRxdP}), we obtain 
\begin{equation}
i\hbar\frac{\partial x}{\partial p}+x^{2}=\frac{2}{k}
\left[\tilde{E}-\left(\frac{p^{2}}{2m}-\frac{p^{4}}{8m^{3}c^{2}}\right)\right].
\label{qRxdPcoordeqn}
\end{equation}
 We impose the physical boundary condition that in the limit $\hbar \rightarrow 0$,
the quantum coordinate function $x(p,E)$ should reduce to the classical coordinate function
$x_{C}(p,E)$ for all $p$. This equation has the same structure as its 
non-relativistic counterpart, Eq. ~(\ref{xpquantumorbit}).

Following the classical case, we define the quantum action variable
as $J_{WR}=-\frac{1}{2\pi}\oint_{C_{WR}'}x_{WR}(p,E)dp$, where $x_{WR}$
is the solution of Eq.~(\ref{qRxdPcoordeqn}) that satisfies the physical
boundary condition. The construction of this form of the quantum action variable and 
the derivation of energy eigenvalues in this case is identical in all relevant details
to the problem of the the non-relativistic anharmonic oscillator with
a quartic potential energy term $\delta x^{4}$, and with $J$ in
the $\oint p(x,E)dx$ form, shown in the Appendix. A comparison with
the latter problem indicates that we can find the action variable
for the weakly relativistic oscillator by making the replacements
$x\rightarrow-p, p\rightarrow-x, \frac{\delta}{k^{2}}\rightarrow\frac{-1}{8mc^{2}}$
in Eq.~(\ref{jqahoeval}) and get 
\begin{equation}
J_{WR}(\tilde{E})=\frac{\tilde{E}}{\omega_{0}}\;-\;\frac{\hbar}{2}\;+\;\frac{3\hbar}{64}\;
\left\{ 1+\frac{4{\tilde{E}}^{2}}{(\hbar\omega_{0})^{2}}\right\} \;
\left(\frac{\hbar\omega_{0}}{mc^{2}}\right),
\label{jwrquantxdpeval}
\end{equation}
 correct to the first order in $\frac{\hbar\omega_{0}}{mc^{2}}$.
Solving for $\tilde{E}$ in this order using $J_{WR}(E)=n\hbar$,
we get
\begin{equation}
\tilde{E}=\left[\left(n+\frac{1}{2}\right)-\frac{3}{16}\left(\frac{\hbar\omega_{0}}
{mc^{2}}\right)\left\{ \left(n+\frac{1}{2}\right)^{2}+4\right\} \right]\hbar\omega_{0}.
\label{rellevelsxdp}
\end{equation}

The above energy eigenvalues obtained by approximating the quantum
coordinate function, while not identical to those
in Eq.~(\ref{wrpdxlevels}), display the same features that were
observed earlier. Both forms of the quantum action variable indicate that the 
lowering of energy levels due to the weak relativistic correction is 
approximately proportional to $n^2$.

\section{OSCILLATOR FREQUENCIES AND ENERGY LEVELS FROM DIFFERENT SCHEMES: A
COMPARISON}

Table 1 summarizes the expressions for the classical action variable
of the relativistic oscillator under four different schemes, two of which we have
considered here. The first two rows show two equivalent series representations
for $J_{C}$ for the fully relativistic oscillator with no approximation\cite{classreloscfreq}, and the next two
rows the corresponding representations obtained by using the weak
relativistic approximation for the kinetic energy. To the first order
in $\epsilon$ they all yield the same expression. Relativistic dynamics
lowers the oscillator's frequency due to time dilation, and the fraction by which the 
frequency is lowered is $\frac{3}{8}\epsilon$ for this weakly relativistic
oscillator. The quantum action variable for the weakly relativistic oscillator
under the two alternate action variable schemes we considered 
is shown in Table 2. The energy eigenvalues obtained from these schemes
and those from two other approximation schemes are also shown in this
table. The semiclassical JWKB approximation involves discretization
of the \textit{classical} action variable $J_{CWR}(\tilde{E})$ in
Eq. ~(\ref{weakjcrpdx}) for the weakly relativistic oscillator:
\begin{eqnarray*}
J_{CWR}(\tilde{E})\;=\;\frac{\tilde{E}}{\omega_{0}}\left[1+\frac{3}{16}\epsilon\right]\;
=\;(n+\frac{1}{2})\:\hbar.
\end{eqnarray*}
 In the Rayleigh-Schrödinger perturbation scheme the shifts in energy
eigenvalues from their unperturbed values, to the first order in $\epsilon$,
are $\Delta\tilde{E}_{n}=<\phi_{n}|\hat{H_{\epsilon}}|\phi_{n}>$,
treating $\hat{H_{\epsilon}}=-\frac{p^{4}}{8m^{3}c^{2}}$ as the perturbation
term in the Hamiltonian whose zeroth order form is $\hat{H_{0}}=\frac{{\hat{p}}^{2}}{2m}+\frac{1}{2}k{\hat{x}}^{2}$.
In all four schemes the calculated weak relativistic shift in the
energy eigenvalues from thier non-relativistic values is approximately
$-\frac{3}{16}\left(\frac{\hbar{\omega}_0}{mc^2}\right)n^2$.
We find that the energy level spacing is 
\begin{eqnarray*}
E_{n+1}-E_{n}\;\approx\;\hbar\omega_{0}\left[1-\frac{3}{8}n\:\frac{\hbar\omega_{0}}{mc^{2}}
\right]\;\approx\;\hbar\omega_{0}\left[1-\frac{3}{8}\:\epsilon\right]\;=\;\hbar\omega_{WR}.
\end{eqnarray*}
 The angular frequency $\omega_{WR}$, unlike in the non-relativistic
case, is energy dependent. Thus the spacing between the quantum energy
levels in a range which conform to this approximation scheme is proportional
to the classical oscillator's weak relativistic angular frequency
for that energy range. Viewed differently, the weak relativistic correction produces a fractional
shift of $\frac{3}{8}\epsilon$ in the angular frequency of the classical oscillator. There is
an identical fractional shift in the energy level separation of the corresponding quantum oscillator.

\section{QUANTUM - CLASSICAL AND RELATIVISTIC - NON-RELATIVISTIC CORRESPONDENCE}

The physically correct theory of matter should be both quantum in nature
and meet the principle of relativity. The Klein-Gordon and Dirac
equations represent physical theories that meet these two physical
requirements. The description of the weakly relativistic oscillator
considered here meets the first principle. While it does not meet
the second principle it does incorporate a weak relativistic dynamical
correction. It is a useful treatment of an \char`\"{}energetic\char`\"{}
oscillator as it begins to approach the relativistic regime. We notice
the operation of two correspondence principles in the dynamics
of this physical system. One is the reduction of the relativistic
model to its appropriate, well established non-relativistic form.
The other is the similar reduction of the quantum model to its corresponding
classical form. Both the weak relativistic quantum Hamilton-Jacobi
equations ~(\ref{HJWpdx}) and ~(\ref{HJweakqRxdP}) reduce to their
weak relativistic \textit{classical} counterparts in the limit $\hbar\rightarrow0$,
and these equations, in turn, assume their classical non-relativistic
form for $\epsilon << 1$. Secondly, the poles of the quantum momentum function $p_{WR}(x,\tilde{E})$
in the complex-$x$ plane, and those of the quantum coordinate function
$x_{WR}(p,\tilde{E})$ in the complex-$p$ plane coalesce to form
the branch cuts of classical functions $p_{CWR}(x,\tilde{E})$ and
$x_{CWR}(p,\tilde{E})$ respectively. Thirdly, the expressions for
$J_{WR}(\tilde{E})$ in the quantum case reduce to those of $J_{CWR}(\tilde{E})$
in the classical non-relativistic limit. Fourthly, the weakly relativistic
quantum action variable $J_{WR}(\tilde{E})$ reduces to the non-relativistic
quantum action variable $J(\tilde{E})$ for $\epsilon << 1$. This formalism
of quantum mechanics, applied to the harmonic oscillator, thus shows in a unified framework
the critical role played by the two \char`\"{}small\char`\"{}
parameters, $\hbar$ and $\epsilon$, in the emergence of non-relativistic
classical theory from non-relativistic quantum theory and a non-relativistic
quantum model from a weakly relativistic quantum model.

\section{CONCLUSION}

The use of action-angle variables in the study of classical periodic
systems has a parallel in quantum mechanics. Using the quantum action
variable the exact energy eigenvalues of a bound system that is classically
periodic can be obtained without a detailed solution of the dynamical
equations. We have demonstrated the construction of two forms of the
quantum action variable, and applied it to the harmonic oscillator
to obtain its energy levels. We have extended the use of this formalism
to a weakly relativistic harmonic oscillator. The classical frequency
of such an oscillator is lowered from its non-relativistic value by
the fraction $\frac{3}{8}\epsilon$. There is an identical fractional
shift in the energy level separation of the corresponding quantum
harmonic oscillator due to the leading order relativistic correction.
While the problem must be properly addressed by a relativistic quantum
theory we have shown how this system may be studied in a unified manner
within a non-relativistic quantum Hamilton-Jacobi theory.

\section{APPENDIX: ANHARMONIC OSCILLATOR WITH QUARTIC TERM}

We consider an approximation scheme for obtaining the energy eigenvalues
for a quartic anharmonic oscillator described by the potential $V(x)=\frac{1}{2}kx^{2}+\delta x^{4}$,
with $\delta>0$, using the quantum action variable. The $\delta<0$
case can be treated in a similar manner. The term $\delta x^{4}$
is considered \char`\"{}small\char`\"{} for $|x|<{x_{2}}^{(0)}$ in
comparison to the dominant term $\frac{1}{2}kx^{2}$. Dynamical variables specific
to this oscillator have the suffix $AHO$. We denote the physical
turning points for the $\delta=0$ (or the simple harmonic oscillator)
case by $-x_{1}^{(0)}=x_{2}^{(0)}=\sqrt{\frac{2E}{k}}$. 
$x_{1}$ and $x_{2}$ are the physical turning points in the presence of the
quartic potential term, and are approximately of magnitude 
$x_{2}^{(0)}\left( 1-\frac{2E}{k^2}\delta \right )$. For $\delta>0$ they are closer to
each other than those in the harmonic oscillator case.  We seek
the energy eigenvalues of this anharmonic oscillator correct to order $\delta^{1}$.
The classical momentum function $p_{CAHO}(x,E)$ has two other (unphysical)
turning points, approximately given by $-x_{3}=x_{4}\approx i\;\sqrt{\frac{k}{2\delta}} 
\left( 1+\frac{2E\delta}{k^2} \right)$.
We write the classical momentum function in the form 
\begin{equation}
p_{CAHO}(x,E)=i\sqrt{mk}\; x\;\left[1-\left(\frac{x_{2}^{(0)}}{x}\right)^{2}\right]^{\frac{1}{2}}
\;\left[1+\frac{\frac{2\delta}{k}x^{2}}{1-\left(\frac{x_{2}^{(0)}}{x}\right)^{2}}\right]^{\frac{1}{2}}.
\label{pcaho}
\end{equation}
 To order $\delta^{1}$ this has, for $x_{2}^{(0)}<|x|<|x_{4}|$,
the series representation 
\begin{equation}
p_{CAHO}(x,E)\approx i\sqrt{mk}\; x\;\left[\sum_{j=0}^{\infty}c_{j}
\left(\frac{x_{2}^{(0)}}{x}\right)^{2j}\right]\;\left[1+\frac{\delta}{k}x^{2}\;
\left\{\sum_{l=0}^{\infty}\left(\frac{x_{2}^{(0)}}{x}\right)^{2l}\right\}\right].
\label{pcahoseries}
\end{equation}
 Here $c_{j}$ are the coefficients in the binomial expansion of $\sqrt{1-u}$
in powers of $u$. We use a similar structure for the Laurent series, 
for the anharmonic oscillator's quantum momentum function $p_{AHO}(x,E)$,
good for $x_{2}<|x|<|x_{4}|$ and correct to order $\delta^{1}$: 
 \begin{equation}
p_{AHO}(x,E) \approx i\sqrt{mk}\; x\;\left[\sum_{j=0}^{\infty}b_{j}
\left(\frac{x_{2}^{(0)}}{x}\right)^{2j}\right]\;
\left[1+\frac{\delta}{k}x^{2}\;\left\{\sum_{l=0}^{\infty}D_{l}\;
\left(\frac{x_{2}^{(0)}}{x}\right)^{2l}\right\}\right]
\label{pquantahoseries}
\end{equation}
The coefficients $b_{j}$ are known from our solution of the quantum
harmonic oscillator problem (Section IV.B). We need to solve for
the coefficients $D_{l}$ using Eq.~(\ref{peq}). The quantum action
variable is defined as 
\begin{equation}
J_{AHO}(E)=\frac{1}{2\pi}\oint_{C_{AHO}}p_{AHO}(x,E)dx,
\label{jqahodef}
\end{equation}
where the counterclockwise rectangular contour $C_{AHO}$ wraps around the real axis between the
turning points $x_{1}$ and $x_{2}$ (or $x_{1}^{(0)}$
and $x_{2}^{(0)}$ for the $\delta<0$ case). The coefficient of $x^{-1}$ in the series in
Eq.~(\ref{pquantahoseries}) is $im\omega_{0}{x_{2}^{(0)}}^{2}\;
\left[b_{1}+\frac{\delta}{k}{x_{2}^{(0)}}^{2}\;(b_{2}D_{0}+b_{0}D_{2}+b_{1}D_{1})\right]$.

We define the dimensionless parameter $\lambda=\frac{\hbar\omega_{0}}{4E}$,
which is "small" for large values of the oscillator's energy.
Introducing the above series for $p_{AHO}(x,E)$ in Eq.~(\ref{pquantahoseries})
and solving for the D-coefficients, we get $D_{0}=1,\; D_{1}=1+\lambda,\; D_{2}=1-\frac{3}{2}\lambda+2\lambda^{2}$.

Evaluating $J_{AHO}$ by the outward deformation of the contour $C_{AHO}$
into the circular counterclockwise contour $\gamma_{AHO}$ we obtain 
\begin{equation}
J_{AHO}=\hbar\left[-\frac{1}{2}+\frac{1}{4\lambda}-\frac{3\delta\hbar}
{32m^{2}\omega_{0}^{3}}\;\left(4+\frac{1}{\lambda^{2}}\right)\right].
\label{jqahoeval}
\end{equation}
Solving for $E$ in $J_{AHO}(E)=n\hbar$ (obtained from inwardly deforming
$C_{AHO}$ to encircle the poles and integrating) we get the energy
eigenvalues of the quartic anharmonic oscillator, to order $\delta^{1}$:
\begin{equation}
E=\hbar\omega_{0}\;\left[\left\{ n+\frac{1}{2}+\frac{3}{8}
\left(\frac{\delta}{k^{2}}\right)\;\hbar\omega_{0}\right\} 
+\frac{3}{2}\left(\frac{\delta}{k^{2}}\right)\;\hbar\omega_{0}\;
\left\{ n+\frac{1}{2}+\frac{3}{8}\left(\frac{\delta}{k^{2}}\right)\;
\hbar\omega_{0}\right\} ^{2}\right]
\label{ahoenergylevels}
\end{equation}
For $\delta>0$, the quartic perturbation shifts
the oscillator's energy levels higher (and lower for the $\delta<0$
case), with the spacing between the energy levels increasing linearly
with $n$.

\newpage{}

\section*{Tables}

\begin{table}[h]

\caption{Action variable for a classical relativistic harmonic oscillator from 4 schemes.}

\begin{tabular}{lll}
\hline
{\bf Nature of oscillator} &
{\bf Form of $J_{C}$ }&
{\bf Expression for $J_{C}(\tilde{E})$ }\tabularnewline \hline
\tableline

Fully relativistic case &
$\frac{1}{2\pi}\oint p dx$ &
$\frac{\tilde{E}}{{\omega}_{0}}\;\sqrt{(1+\frac{\epsilon}{2})}\;\left[1-\frac{1}{8}\left(\frac{\epsilon}{2+\epsilon}\right)-\frac{1}{64}\left(\frac{\epsilon}{2+\epsilon}\right)^{2}\ldots\right]$ \tabularnewline
\tableline

Fully relativistic case &
$-\frac{1}{2\pi}\oint x dp$ &
$\frac{\tilde{E}}{{\omega}_{0}}\;\sqrt{1+\frac{\epsilon}{2}}\;\left[1-\frac{1}{16}\epsilon+\frac{7}{256}{\epsilon}^{2}+\frac{1}{128}{\epsilon}^{3}\ldots\right]$ \tabularnewline
\tableline

Weak relativistic case &
$\frac{1}{2\pi}\oint p dx$ &
$\frac{\tilde{E}}{{\omega}_{0}}\;(1+\frac{3}{16}\epsilon)$, to order
$\epsilon$ \tabularnewline
\tableline

Weak relativistic case &
$-\frac{1}{2\pi}\oint x dp$ &
$\frac{\tilde{E}}{{\omega}_{0}}\;(1+\frac{3}{16}\epsilon)$, to order
$\epsilon$ \tabularnewline
\hline
\end{tabular}
\label{table1} 
\end{table}

\begin{table}[h]

\caption{Correction to zeroth order energy eigenvalues, $\tilde{E}_{n,0}=(n+\frac{1}{2})\:\hbar\omega_{0}$,
of weakly relativistic harmonic oscillator.}

\begin{tabular}{lll}
\hline
{\bf Calculation scheme } &
{\bf Calculated term } &
{\bf Correction to $\tilde{E}_{n,0}$ } 
\tabularnewline
\hline 
Evaluate quantum action   &
$J_{WR}=\frac{\tilde{E}}{{\omega}_{0}}\;[1+\epsilon\{\frac{3}{16}+$ &
$-\frac{3}{16}\hbar{\omega}_{0}\left[(n+\frac{5}{3})^{2}-(\frac{5}{3})^{2}\right]\frac{\hbar{\omega}_{0}}{mc^{2}}$ \tabularnewline
variable in $\frac{1}{2\pi}\oint p dx$ form &
$\frac{7}{16}\left(\frac{\hbar{\omega}_{0}}{\tilde{E}}\right)-\frac{17}{64}\left(\frac{\hbar{\omega}_{0}}{\tilde{E}}\right)^{2}\}]-\frac{\hbar}{2}$ &
\tabularnewline
\hline 
Evaluate quantum action   &
$J_{WR}=\frac{\tilde{E}}{\omega_{0}}-\frac{\hbar}{2}$ &
$-\frac{3}{16}\hbar\omega_{0}\left[\left\{ n+\frac{1}{2}-\frac{3}{4}(\frac{\hbar\omega_{0}}{mc^{2}})\right\}^{2}+4\right]
\frac{\hbar\omega_{0}}{mc^{2}}$ \tabularnewline
variable in $-\frac{1}{2\pi}\oint x dp$ form &
$+\frac{3\hbar}{64}\left\{ 1+\frac{4{\tilde{E}}^{2}}{(\hbar\omega_{0})^{2}}\right\} 
\frac{\hbar\omega_{0}}{mc^{2}}$
 &
\tabularnewline
\hline 
JWKB approximation &
$J_{CWR}=\frac{\tilde{E}}{\omega_{0}}\left[1+\frac{3}{16}\epsilon\right]$ &
$-\frac{3}{16}\hbar\omega_{0}(n+\frac{1}{2})^{2}\;\frac{\hbar\omega_{0}}{mc^{2}}$ 
\tabularnewline
\hline
Rayleigh-Schrödinger &
$\Delta{E_{n}}=<\phi_{n}|-\frac{{\hat{p}}^{4}}{8m^{3}c^{2}}|\phi_{n}>$ &
$-\frac{3}{16}\hbar\omega_{0}[(n+\frac{1}{2})^{2}+\frac{1}{4}]\;\frac{\hbar\omega_{0}}{mc^{2}}$ \tabularnewline
perturbation theory &
&
\tabularnewline
\hline
\end{tabular}
\label{table2} 
\end{table}

\end{document}